\documentstyle[epsfig]{aipproc}
\newcommand{\cL}{{\cal L}}

\newcommand{\prd}{{\em Phys.\ Rev.\ }  {\bf D}}

\newcommand{\prl}{{\em Phys.\ Rev.\ Lett.\ }}
\newcommand{\np}{{\em Nucl.\ Phys.\ }{\bf B}}
\newcommand{\pl}{{\em Phys.\ Lett.\ }{\bf B}}

\newcommand{\rmp}{{\em Rev.\ Mod.\ Phys.\ }}

\begin{document}
\title{Theoretical Issues in \\ Rare $K$, $D$ and $B$ Decays\footnote{Talk 
presented at the Workshop on Heavy Quarks at Fixed 
Target, Fermilab, Batavia IL, October 10-12, 1998.} }

\author{Gustavo Burdman\thanks{\ttfamily burdman@pheno.physics.wisc.edu}}
\address{
University of Wisconsin, Madison, WI 53706, USA.
}

\maketitle

\setlength{\unitlength}{1cm}
\begin{picture}(0,0)(0,0)
\put(11.5,6){MADPH-98-1093}
\end{picture}
\vspace{-24pt}

\begin{abstract}
I discuss some theoretical aspects of rare $K$, $D$ and $B$ decays
focusing mainly on their potential as tests of the one loop structure
of the standard model. I concentrate on flavor changing 
neutral current processes 
and compare our ability to extract short distance 
physics in the three cases. Finally, I give some examples of the 
sensitivity of these decays to extensions of the standard model.
\end{abstract}

\section*{Introduction}

The continuing success of the standard model (SM) has turned into a challenge
both for theorists and experimentalists alike. On the one hand, theorists 
believe that the SM picture of the Higgs mechanism based on one elementary
scalar doublet is unnatural, trivial
and has come to be viewed as an effective description of a more complicated 
Higgs sector, one involving perhaps additional scalars 
and/or fermions or even new gauge interactions. In addition, the idea that  
fermion masses arise as a result of the interactions with this elementary 
scalar, requires for instance that its dimensionless couplings to two 
otherwise 
identical fermions such as the up and the top quarks, differ by more than 
four orders of magnitude. Understanding the mechanism for electroweak 
symmetry breaking (EWSB) and the origin of fermion masses calls for 
physics beyond the SM. At least in the case of EWSB, it is understood 
that this new physics must reside at an energy scale not far beyond
$1~$TeV. The experimental challenge of finding  new physics in direct 
searches may still take some time if the new states or their effects 
only set in at several hundred GeV. A complement of these direct signals 
at the highest available energies is the measurement of the effects 
of the new particles in loops, either through precision measurements 
such as the ones performed at LEP, or through the detection of processes
only occurring at one loop in the SM. Among these are the transitions 
induced by flavor changing neutral currents (FCNC), such as $K^0-\bar{K^0}$
or $b\to s\gamma$. These are forbidden at tree level in the SM due to the 
presence of the Glashow-Illiopoulos-Maiani (GIM) mechanism. 
As a consequence, these processes are highly suppressed in the SM. 
Thus the one loop effect of a new heavy state may translate into a 
large effect in a branching ratio if this loop induces a FCNC transition. 
Here I discuss the FCNC decays of $K$, $D$ and $B$ and their potential 
as tools in searching for new physics at high energy scales $\gtrsim M_W$.
One critical aspect in evaluating this potential is the extent to which 
a given process is determined by high energy scales, i.e. short distance
physics, or it is contaminated by the more mundane effects of long 
distance dynamics such as the ones induced by propagating intermediate
hadrons. Since the latter are not calculable in perturbation theory,  
the decay modes affected by long distance physics are less 
understood. 
Next, I will give examples of short and long distance contributions 
in the FCNC decays of $K$, $D$ and $B$ mesons. 
Once established which modes are the most likely to be testing grounds of the 
SM, I will turn to the discussion of two typical examples of its extensions: 
supersymmetry and the effects of anomalous triple gauge boson couplings. 

\section*{Rare $K$ Decays} 
Let us start the discussion with the decay $K^+\to\pi^+e^+e^-$. 
In the SM this  process receives one loop contributions 
from electroweak penguin and box diagrams involving up-type quarks.
These are truly short distance diagrams and although
some long distance physics enters in the hadronization from 
$s\to K$ and from $d\to \pi$ in the way of form-factors, the 
theoretical uncertainties this introduces are not by themselves 
large enough to obscure the interesting physics. 
However, the process $K^+\to\pi^+\gamma^*$ followed by 
$\gamma^*\to e^+e^-$ gives another contribution, 
reflecting  the long distance dynamics of the 
$K^+\pi^+\gamma$ vertex, and dominates the rate~\cite{valencia}. 
We conclude that this decay 
mode is not well suited for a test of the short distance physics entering the 
loops. 

We now turn to consider $K_L\to\pi^0e^+e^-$. The short distance 
contributions to this process involve direct CP violation. 
This is extremely interesting since the measurement of this decay rate 
could be 
in principle a direct determination of the CP violating parameter $\eta$, 
the complex phase in the Cabibbo-Kobayashi-Maskawa (CKM) matrix. 
Furthermore, the one photon intermediate state cannot contribute here since 
the long distance vertex to one photon is CP conserving. 
Still, there is contamination from long distance 
physics. This  comes from two sources. First, there is indirect CP violation
given essentially by 
\begin{equation}
BR(K_L\to\pi^0e^+e^-)_{\rm ICP} = |\epsilon_K|^2\frac{\tau(K_L)}{\tau(K_S)}
BR(K_S\to\pi^0e^+e^-),
\label{icp}
\end{equation}
where the $K_S$ decay, being CP conserving, is dominated by the 
one photon intermediate state. In addition, there is a CP conserving 
contribution to the decay rate from $K_L\to\pi\gamma^*\gamma^*$ which 
gives a non interfering term resulting in 
$BR(K_L\to\pi^0e^+e^-)_{\rm CPC}\simeq (1-2)\times 10^{-12}$, although 
this could be larger~\cite{donoghue}.
The indirect CP violating piece will be well known once the $K_S$ decay mode
is measured. Progress to a better understanding of the CP conserving piece 
may be achieved in the near future. Thus, there is reasonable hope
that in this process we may be able to disentangle the short and long distance 
physics, at least to some extent. For the moment, it remains in the list of 
long distance ``polluted'' modes. 

Finally, we turn to the decay modes with neutrinos replacing the charged leptons in the
final state. The processes  $K^+\to\pi^+\nu\bar\nu$ and $K_L\to\pi^0\nu\bar\nu$ are 
almost completely determined by short distance physics. The effective 
Hamiltonian for the charged mode is 
\begin{equation}
{\cal H}_{\rm eff}=\frac{4G_F}{\sqrt{2}}\frac{\alpha}{2\pi s^2\theta_W}
\sum_\ell\left(\lambda_cX^\ell_c + \lambda_tX(x_t)\right)
(\bar s_L\gamma_\mu d_L)(\bar\nu_L\gamma^\mu\nu_L)_\ell,
\label{heff}
\end{equation}
where $\lambda_i=V_{is}^*V_{id}$, 
$(\ell=e,\mu,\tau)$, $X(x_t)$ is the result of the top quark loop contribution,
and $X^\ell_c$ is the charm quark contribution and carries a dependence on the lepton 
flavor coming from the box diagram with charged leptons. The contribution from the 
charm loop, in addition to the charm quark mass dependence, 
introduces a sizeable scale dependence which is only reduced when including
next-to-leading-order corrections. The remnant dependence on the scale $m_c$ entering in 
the charm running mass results in an uncertainty $<10\%$ in the  $BR(K^+\to\pi^+\nu\bar\nu)$.
The hadronic matrix element needed to compute the matrix element of the exclusive mode
can be obtained by isospin rotation from $K^+\to\pi^0\ell^+\nu$ and therefore does not 
introduce additional hadronic uncertainties.
The SM prediction for the branching ratio is obtained for  $m_c=1.3~$GeV, $m_t=170~$GeV, 
$V_{cb}=0.04$ and $V_{ub}=0.032$, leading to~\cite{buras}
$Br(K^+\to\pi^+\nu\bar\nu)=(9.1\pm3.8)\times 10^{-11}$,
where the uncertainty is mainly from the CKM parameters.
The recent observation of one event in this channel by BNL E787, translates into~\cite{bnl} 
$
Br(K^+\to\pi^+\nu\bar\nu)_{\rm exp}=(4.2+9.7-3.5)\times 10^{-10}$. 

Finally, the neutral mode  $K_L\to\pi^0\nu\bar\nu$ is even cleaner than the charged 
one due to the fact that it is largely dominated by direct CP violation. 
As a result, only the top quark loop in Eq.~(\ref{heff}) contributes so the
uncertainties associated with the charm scale are not present. 
Furthermore, this mode constitutes a very unique and direct way to measure the 
CP violating phase of the CKM matrix.
The amplitude depends on $Im[\lambda_t]$, which in the Wolfenstein parametrization can be 
written in term of the CP violating term $\eta$ and $V_{cb}$, giving~\cite{buras}
\begin{equation}
Br(K_L\to\pi^0\nu\bar\nu)=3.29\times10^{-5}\eta^2 |V_{cb}|^4 X^2(x_t),
\label{neutral}
\end{equation}
with $\eta\equiv Im[V_{ub}^*/V_{cb}]/\lambda$.
The branching ratio of the neutral mode is still too small in the SM 
when compared with current experimental limits. 
Future experiments expect to be sensitive to a few $\times 10^{-11}$ branching fraction. 

There seems to be a compromise between experimental accessibility and the theoretical 
uncertainties in any of the modes discussed above: the most accessible modes tend to be 
affected by larger theoretical uncertainties. This tension is always present in 
rare FCNC decays. In the case of $K$ decays, the neutrino modes are hard but still 
accessible experimentally. The charged kaon mode seems to be a good compromise, since is 
a short distance dominated process and the uncertainties seem to be surmountable.
The additional interest of the neutral mode is the observation of direct CP violation 
and the direct measurement of CKM parameters. 

\section*{Rare $D$ {\mbox vs.} $B$ Decays}

We now turn to a comparative discussion of FCNC in $D$ and $B$ decays. 
The essential aspects can be framed as external-up-quark 
vs. external-down-quark  processes. 
We will make the comparison using the radiative processes
$c\to u\gamma$ vs. $b\to s\gamma$, for the sake of simplicity. 
Most of the conclusions can be extended to other modes. 
The short distance contributions to the radiative FCNC process $Q\to q\gamma$ 
result in the decay width
\begin{equation}
\Gamma^{(0)}=\frac{\alpha G_F^2}{128\pi^4}m_Q^5\left|\sum_i
\lambda_i F(x_i)\right|^2,
\label{radwid}
\end{equation}
where the superscript ``$0$'' denotes the absence of QCD corrections, $Q=(c,b)$, 
$x_i=(m_{q_i}/M_W)^2$, and the 
function $F(x)$ is the result of integrating the loop contribution 
of the internal quark $i$. This loop function is the same in  the $c$ and $b$ 
cases. The main difference in Eq.~(\ref{radwid}) comes from the masses of the internal quarks
and the CKM factors $\lambda_i$. In order to see how this affects the widths we turn
to Table~I, where we show separately the contribution of each quark flavor.

\phantom{xxxx}\vspace{0.1in}
\begin{center}
\begin{tabular}{|c||c|c|c|}
\hline
 & $i$ & $F(x_i)$ & $\lambda_i F(x_i)$ \\ \hline
 & d & $1.6 \times 10^{-9}$ & $3.4 \times 10^{-10}$\\ 
$c\to u\gamma$ & s & $2.9 \times 10^{-7}$ & $6.3 \times 10^{-8}$ \\
 & b & $ 3.3 \times 10^{-4}$ & $3.2 \times 10^{-8}$ \\  \hline
 & u & $2.3 \times 10^{-9}$ & $1.3 \times 10^{-12}$\\ 
$b\to s\gamma$ & c & $2.0 \times 10^{-4}$ & $7.3 \times 10^{-6}$ \\
 & t & $0.4$ &  $1.6 \times 10^{-2}$ \\  \hline
\end{tabular}
\begin{center}
Table~I:~Contributions to $Q\to q\gamma$. From Ref.~\cite{chrad}. 
\end{center}
\end{center}
\phantom{xxxx}\vspace{0.1in}
The CKM factors are $\lambda_i=V_{ci}^*V_{ui}$ for  $c\to u\gamma$, and 
$\lambda_i=V_{ib}^*V_{is}$ for $b\to s\gamma$. As we can see, all three contributions
are small in the $c\to u\gamma$ case, whereas for $b\to s\gamma$ the top quark loop
gives the overwhelmingly dominant piece. The central point is that heavier quarks give
the dominant contributions as long as their mixing with the external quarks is 
not highly suppressed. This is a consequence of the non-decoupling aspect of the SM, 
the fact that fermions that acquired masses from the Higgs mechanism do not decouple 
in loops involving the massive electroweak gauge bosons. 
In $c\to u\gamma$ the internal $b$ quark contribution would dominate if it was not for 
the fact that $V_{cb}$ and $V_{ub}$ are extremely small. 
In any event, the QCD uncorrected $c\to u\gamma$ rate is very small due to the 
CKM dominance of the lighter intermediate states $d$ and $s$. 
The $b\to s\gamma$ width is large due to the presence of a heavy top!

Although the QCD corrections to Eq.~(\ref{radwid}) are generally important, their 
impact also varies depending on the intermediate mass in the loop. 
They enhance the $c\to u\gamma$ rate by five orders of magnitude on the one hand, 
but the $b\to s\gamma$ rate goes up by less than a factor of three or so. 
The main source of these large corrections is the mixing of the short distance 
operators such as 
\begin{equation}
{\cal O}_7=\frac{e}{16\pi^2} m_Q(\bar q_L\sigma_{\mu\nu}Q_R) F^{\mu\nu},
\label{o7}
\end{equation}
generated by the interesting short distance physics, with the more mundane
four-fermion operators such as 
\begin{equation}
(\bar q_L\gamma_\mu q'_L)(\bar q'_L\gamma^\mu Q_L),   
\label{ffop}
\end{equation}
that are generated at tree level by the SM charged currents.   
The mixing comes about when loop generated by gluons are taken into account. 
New physics, if present,  will almost certainly appear in (\ref{o7}), not in 
(\ref{ffop}), which is then a background for precision tests of the SM. 
Thus, the lesson from Table~II is that in $b\to s\gamma$
there is still sensitivity 
to new physics affecting the operator (\ref{o7}),whereas even if it were 
possible to measure a branching ratio as low as $10^{-12}$ for $c\to u\gamma$, 
this would reflect the SM physics of operators such as the one in Eq.~(\ref{ffop}). 
\phantom{xxxx}\vspace{0.1in}
\begin{center}
\begin{tabular}{|c||c|c|}
\hline
 & No QCD & QCD Corrected  \\ \hline
  $Br(c\to u\gamma)$~\cite{chrad}& $1.5 \times 10^{-17}$ & $6.0 \times 10^{-12}$\\ 
 $Br(b\to s\gamma)$~~~~&  $1.3\times 10^{-4}$ & $3.3 \times 10^{-4}$ \\ \hline
\end{tabular}
\begin{center}
Table~II:~Leading order and QCD-corrected branching ratios for  $Q\to q\gamma$.
The QCD corrected rates involve important QCD uncertainties. 
\end{center}
\end{center}
\phantom{xxxx}\vspace{0.1in}
The important point is that the overwhelming dominance of the QCD corrections in 
$c\to u\gamma$ not only tells us that the short distance physics is not sensitive 
to the one loop FCNC operators of interest, but also signals that there will be 
even larger long distance contributions to the rate. After all, the QCD corrections
were computed perturbatively. In general, the dominance of 
the perturbative one loop amplitude by light quark contributions 
hints the existence of large long distance effects.
Although these cannot be computed from first principles, it is possible to estimate 
them 
phenomenologically. 
For instance, the operator (\ref{ffop}) with $q'=s$ gives rise to the dominant 
short distance piece in $c\to u\gamma$, through a $\bar s s$ loop. 
But one could imagine the $\bar s s$ pair propagating a long distance, 
forming a ``$\phi$'', which turns into a photon via vector meson dominance. 
These and other similar long distance mechanisms~\cite{chrad}
give rise, for instance,  to 
$Br(D^0\to\rho^0\gamma)\simeq 10^{-6}$,
far above the level of the QCD-corrected short distance rates expected for 
$c\to u\gamma$ processes. Many other long-distance dominated radiative $D$ 
decays are at this level. Similar effects are expected in the leptonic modes. 
There, however, the gap between short and long distance physics is less dramatic. 
The inclusive short and long distance branching fractions are~\cite{chlep}, 
respectively, 
\begin{eqnarray}
Br(c\to u\ell^+\ell^-)_{\rm SD}&\simeq &10^{-8}, \\ \label{lepsd}
Br(c\to u\ell^+\ell^-)_{\rm LD}&\simeq &10^{-6}.
\label{lepld}
\end{eqnarray}
Then, although the long distance contributions still dominate in the SM, it is 
still conceivable  that new physics contributions could overcome them.  
For instance, the SM prediction for the exclusive  mode 
$Br(D^0\to\pi^0e^+e^-)\simeq 7\times 10^{-7}$, is still well below the current 
experimental bound, $Br(D^0\to\pi^0e^+e^-)_{\rm exp}<4.5\times 10^{-5}$. 
Some extensions of the SM may give large enhancements in charm processes. 
Although, these effects would be more noticeable in $D^0-\bar{D^0}$ mixing, 
they could also result in $c\to u\ell^+\ell^-$ rates well above $10^{-6}$. 
In general, however, one loop effects from new physics in $D$ decays are 
likely to be small compared to long distance effects. 

On the other hand, the analogous $b$ decays are believed to be dominated by 
short distance physics. For instance, the next-to-leading order SM prediction 
for the 
short distance rate gives~\cite{bsg} 
$Br(b\to s\gamma)=(3.38\pm0.33)\times10^{-4}$. 
Long distance contributions similar to those discussed in radiative charm decays, 
can proceed via the propagation of intermediate $\bar c c$ states, the off-shell 
``$J/\psi$''. Estimates of the pollution due to these states in the inclusive 
rate~\cite{bsgld} cannot be made reliably within controlled approximations. 
They tend to vary 
from one calculation to the next and can be as large as $20\%$.
The current experimental measurements give~\cite{bsgexp}
\begin{eqnarray} 
Br(B\to X_s\gamma)_{\rm CLEO}&=&(3.15\pm0.35\pm0.32\pm0.26)\times10^{-4},\\ 
\label{bsgcleo}
Br(B\to X_s\gamma)_{\rm ALEPH}&=&(3.11\pm0.80\pm0.72)\times10^{-4}.
\label{bsgaleph}
\end{eqnarray}
Thus, for the moment the potential long distance pollution is not problematic, 
but it should be taken into account in the future when precise enough measurements
become available. 

Similar considerations apply to the dilepton modes $b\to s\ell^+\ell^-$. 
In this case the long distance pollution comes in the form of ``spill over'' 
of the $J/\psi$ and $\psi'$ resonant peaks into the continuum~\cite{buch}. 
But in principle
these modes are short distance dominated and together with $b\to s\gamma$ constitute
a stringent test of the SM. In addition to the dipole moment operator 
in (\ref{o7}), 
these modes receive contributions from the operators
\begin{eqnarray}
{\cal O}_9&=&\frac{e^2}{16\pi^2}(\bar s_L\gamma_\mu b_L)(\bar\ell\gamma^\mu\ell),
\label{o9}\\
{\cal O}_{10}&=&\frac{e^2}{16\pi^2}(\bar s_L\gamma_\mu b_L)(\bar\ell\gamma^\mu
\gamma_5\ell).
\label{o10}
\end{eqnarray}
These receive contributions from $Z$ penguin and box diagrams. 
As a result, there are three quantities to be measured: $C_7(m_b)$, $C_9(m_b)$ 
and $C_{10}(m_b)$, the Wilson coefficients evaluated at the relevant
experimental scale.  New physics effects enter as additions to the values of the
coefficients at the high energy scale $E>M_W$. 
The extraction of these quantities is a research program involving 
the inclusive rates $B\to X_s\gamma$, $B\to X_s\ell^+\ell^-$, as well 
as exclusive modes such as $B\to K^{(*)}\ell^+\ell^-$ among others. 
A lot of theoretical effort has gone into understanding the inclusive 
decay rates~\cite{inc}, and the theoretical predictions are under control. 
On the other hand, the exclusive modes are, in principle, affected by 
large theoretical uncertainties due to our poor knowledge of the non-perturbative 
dynamics determining form-factors. Some sound theoretical predictions can be made
based on symmetries~\cite{symm}. In some cases~\cite{aszero}, 
this is enough to extract the 
short distance physics. In any event, in the future all these form-factors will
be obtained from first principle calculations on the lattice~\cite{andreas}, 
where a lot of progress has been made recently in computing weak matrix 
elements~\cite{b2pi}. 

\section*{Sensitivity to New Physics} 
Here we discuss two typical examples of extensions of the SM of very different 
kind: supersymmetry and anomalous triple gauge boson couplings (TGC). 

\subsection*{Supersymmetry}
Supersymmetry is perhaps one of the most popular extensions of the SM.  
However, as many other extensions, it has a FCNC problem: in its most general 
form it does not come with an automatic GIM mechanism, and therefore it may generate
large FCNC effects~\cite{miko}. 
Most of the trouble comes from the fact that the diagonalization
of fermion mass matrices does not, in general, diagonalize the squark mass matrices. 
Thus flavor mixing in the sfermion sector ``misaligned'' with the fermions, are a 
potential disaster in general SUSY scenarios.
Even if the sfermion sector is assumed to be diagonal (or aligned), 
there is an additional source of FCNC effects, coming from the charged Higgs and 
chargino-squark contributions, arising from the standard CKM matrix. These effects, 
then 
are expected to be present at most at the SM level, since they depend only on the 
masses of the charged Higgs, the charginos and the third generation squarks.
In any case, some assumption about the sfermion mass matrices is necessary in order to 
accommodate the FCNC constraints. 
The two possibilities are: (i) sfermion mass matrices are diagonal at some high energy 
scale (e.g. $M_{\rm GUT}$) and small off diagonal elements are generated by the 
running down to the electroweak scale; (ii) they are (partially) aligned with the SM 
fermion mass matrices. 

The vast parameter space of SUSY models includes these off diagonal elements, 
the superpartner and  Higgs sector masses and mixings, and 
allows to accommodate the lack of deviations in FCNC processes such as 
$K^0-\bar{K^0}$ mixing, $b\to s\gamma$, etc. However, one can argue that in most cases
the SUSY effects should be ``naturally'' of the order of $(10-20)\%$
 or larger. Such effects, for
instance, in $K$, $D$ and $B$ mixing, or $b\to s$ and $s\to d$ transitions, are 
hard to see at the moment. But larger effects are also possible. 
For example, even satisfying the current $b\to s\gamma$ and mixing bounds, 
we could still see enhancements of ${\cal O} (1)$ in 
$K\to\pi\nu\bar\nu$~\cite{romanino} and 
$b\to s\ell^+\ell^-$~\cite{bslsu}. 

Moreover, it has been recently argued~\cite{isi}
that if large off diagonal sfermion mixings 
are allowed, the next to leading order expansion in theses mixings reveals the 
possibility of even larger effects in the $s\to d Z$ vertex. This would have a large 
impact in decay modes such as $K\to\pi\nu\bar\nu$, where the $Z$ penguin plays 
a dominant role, resulting in enhancements of the branching ratios 
of one order of magnitude or more, depending on the modes. 
This is an interesting possibility and deserves further study, particularly the 
correlation with possible enhancements in $D$ mixing and rare $D$ decays that 
would result from very large mass insertions in the up-squark sector.

\subsection*{Anomalous Triple Gauge Boson Couplings}
We now turn to examine the potential of rare FCNC decays to constrain anomalous
triple gauge boson couplings (TGC). In general, we can assume in a model independent 
way, that extensions of the SM might modify some of the couplings of fermions and/or 
gauge bosons. In particular, the TGC are of interest since they have not been measured
with such precision as some of the fermion couplings. 
We have in mind deviations from the SM values for the couplings of a pair of $W$ 
to a photon or a $Z$. We would expect that the anomalous TGC encode the physics
of some higher energy scale. 
Imposing $CP$ conservation, the most general form of the 
$WWN$ ($N=\gamma, Z$)
couplings can be written as~\cite{dieter}
\newpage
\begin{eqnarray}
\cL_{WWN}&=&g_{WWN}\left\{i\kappa_N W_{\mu}^{\dagger}W_\nu N^{\mu\nu}
+ig_1^N \left(W_{\mu\nu}^{\dagger}W^\mu N^\nu -
W_{\mu\nu}W^{\dagger\mu} N^\nu\right) \right. \nonumber\\
& & \left. 
+g_5^N\epsilon^{\mu\nu\rho\sigma}(W^\dagger_\mu\partial_\rho 
W_\nu 
-W_\mu\partial_\rho W^\dagger_\nu)N_\sigma
+i\frac{\lambda_N}{M_W^2}W^\dagger_{\mu\nu}W^\nu_{~\lambda} 
N^{\nu\lambda}
\right\}~, 
\label{wwnc}
\end{eqnarray}
with the conventional choices being $g_{WW\gamma}=-e$ and 
$g_{WWZ}=-g\cos\theta$. 
Additionally, the are three CP violating Lorentz invariant terms,
resulting in other 6 parameters: $\tilde\kappa_N$ and $\tilde\lambda_N$, obtained
from (\ref{wwnc}) by replacing $N_{\mu\nu}$ by the dual field strength; 
and $g_4^N$ from a term similar to the second one in Eq.~(\ref{wwnc}).  

Gauge invariance implies 
$g_1^\gamma=1$, $g_5^\gamma=g_4^\gamma=0$. 
Then, in principle, there are $11$ 
new free parameters. 
Two CP conserving ($\Delta\kappa_\gamma$, $\lambda_\gamma$) and two 
CP violating ($\Delta\tilde\kappa_\gamma$, $\tilde\lambda_\gamma$)
affecting the $WW\gamma$ couplings; four CP conserving 
($\Delta g_1^Z$,~ $g_5^Z$,~$\lambda_Z$ and $\Delta\kappa_Z$) and three
CP violating ($\tilde\kappa_Z$, ~$\tilde\lambda_Z$ and $ g_4^Z$) shifting the 
$WWZ$ vertex.  

This is a typical problem of this type of approach, where the model independence 
is traded off by a large number of free parameters the sources of which are 
not known. However, simplification is possible, 
when considering rare $B$ and $K$ decays. 
we can neglect the 
contribution of $\Delta\kappa_Z$,~$\lambda_Z$, as well as the three $WWZ$ 
CP violating anomalous TGC,  
since their effects  are suppressed by powers of the 
small external momenta over $m_Z$. This selective sensitivity is an advantage, rather 
than a handicap, when we view these measurements as complement of other ones made at
higher energies and sensitive to all the $Z$ TGC. 

Thus up to this point, we have $6$ coefficients left. 
However, we can ignore $\lambda_\gamma$ and  $\tilde\lambda_\gamma$ 
if we assume that the dynamics producing these non-SM effects 
resides at a scale parametrically larger than the weak scale, say 
$\Lambda\simeq{\cal O}(1)~$TeV. Although this is is not general, 
I believe this is a reasonable scenario, 
since if this was not the case we should take into account the states 
that are present with weak scale masses (e.g. superpartners, 
weakly coupled scalars, etc.) and not integrate them out as we do in 
an effective coupling approach. 
When we accept this, we see that in an effective Lagrangian approach, these  
coefficients can only be generated by next to leading order operators, which can 
be ignored since they are suppressed by $(M_W^2/\Lambda^2)$ with respect to the 
leading order ones\footnote{This corresponds to the so called non-linear realization of 
the EWSB sector. However, it is also possible to imagine a scenario 
where there is a light scalar similar to the SM Higgs, with all other 
new states above the scale $\Lambda$. In these linear realization scenarios, 
the power counting requires the consideration of $\lambda_N$ and $\tilde\lambda_N$
on the same footing with the other anomalous TGC.
}. 
  
\begin{figure} 
\centerline{
\epsfig{file=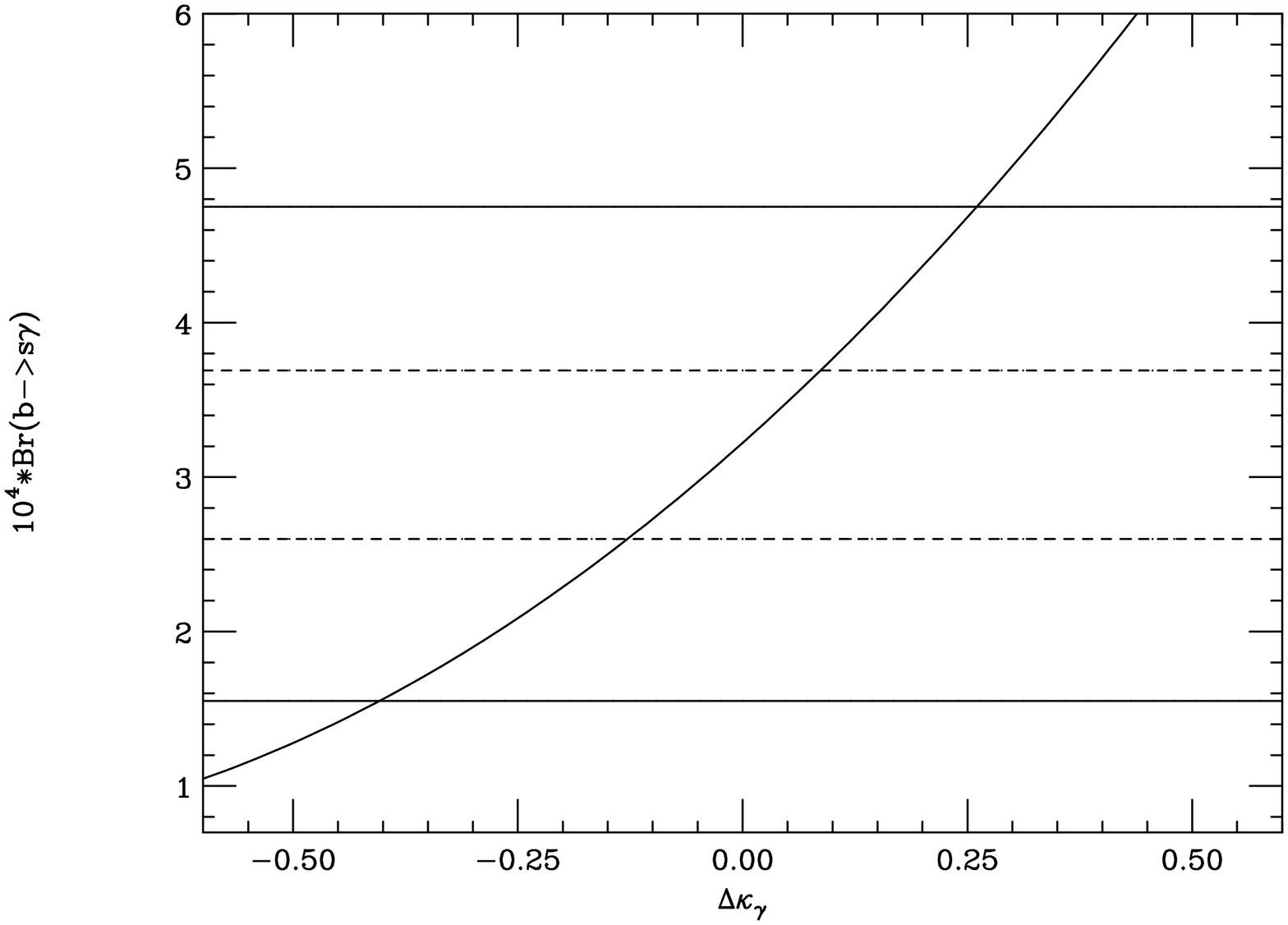,height=2.5in,angle=90}
\epsfig{file=empty.eps,height=0.01in}
\epsfig{file=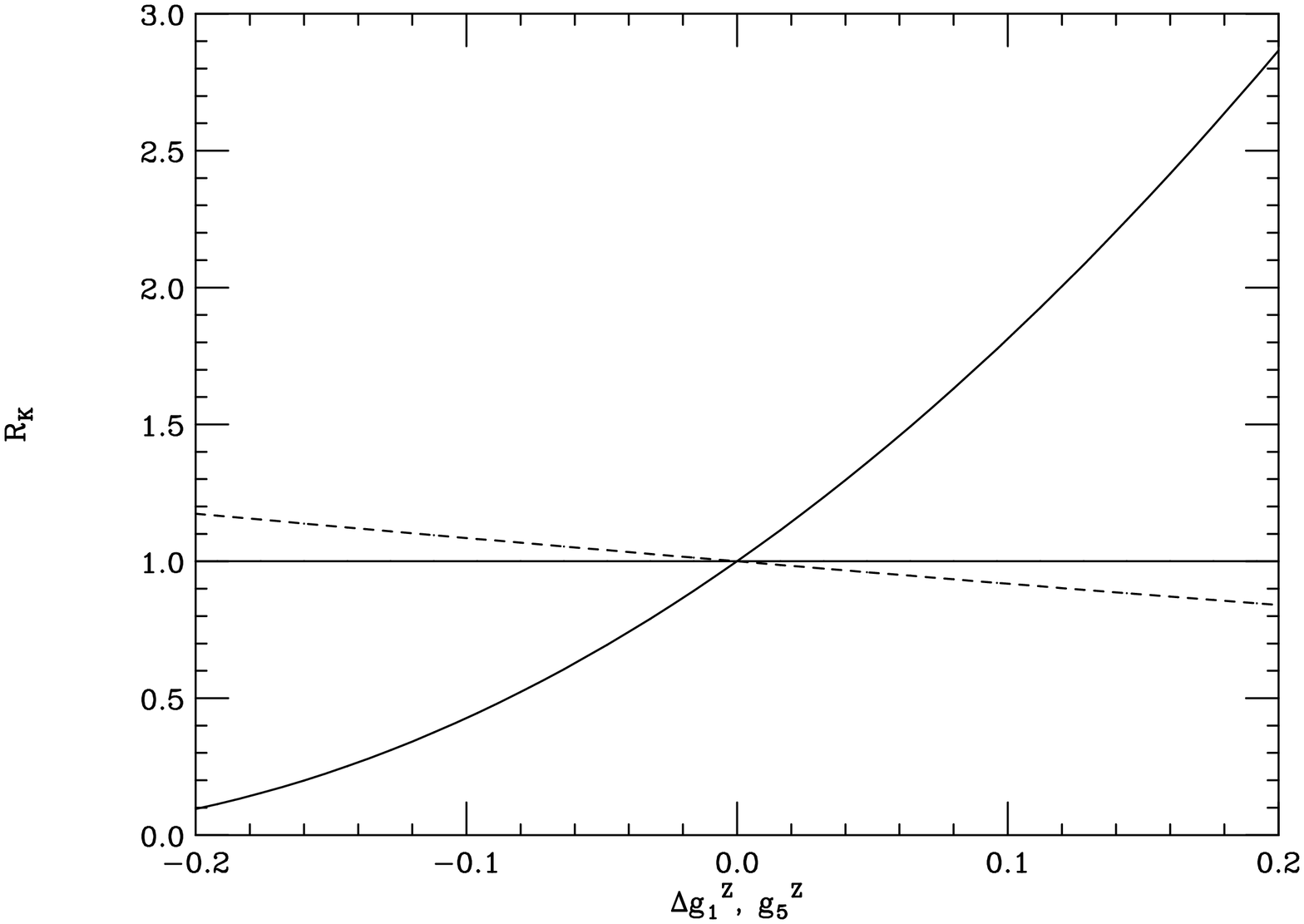,height=2.5in,angle=90}
}
\caption{a) {\em The $Br(b\to s\gamma)$ vs. $\Delta\kappa_\gamma$. 
The solid (dashed) horizontal
lines are the $3(1)\sigma$ CLEO measurement.}
b) {\em The $Br(K^+\to \pi^+\nu\bar\nu)$ normalized to the SM expectation, 
vs. $\Delta g_1^Z$ (solid) and $g_5^Z$ (dashed).
From Ref.~[21].} 
}
\label{fbsg}
\end{figure}

\begin{figure}
\centerline{
\epsfig{file=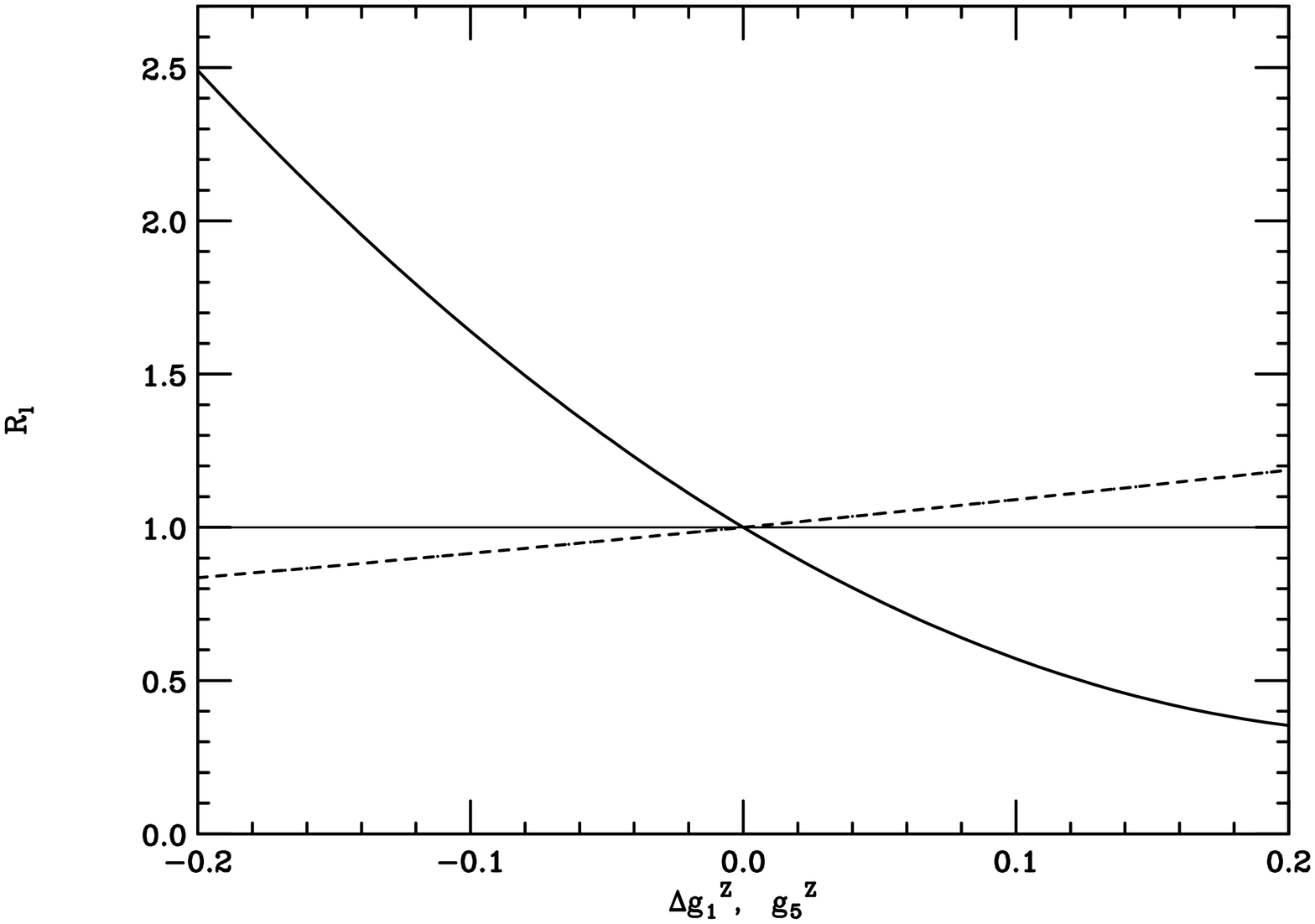,height=2.5in,angle=90}
\epsfig{file=empty.eps,height=0.01in}
\epsfig{file=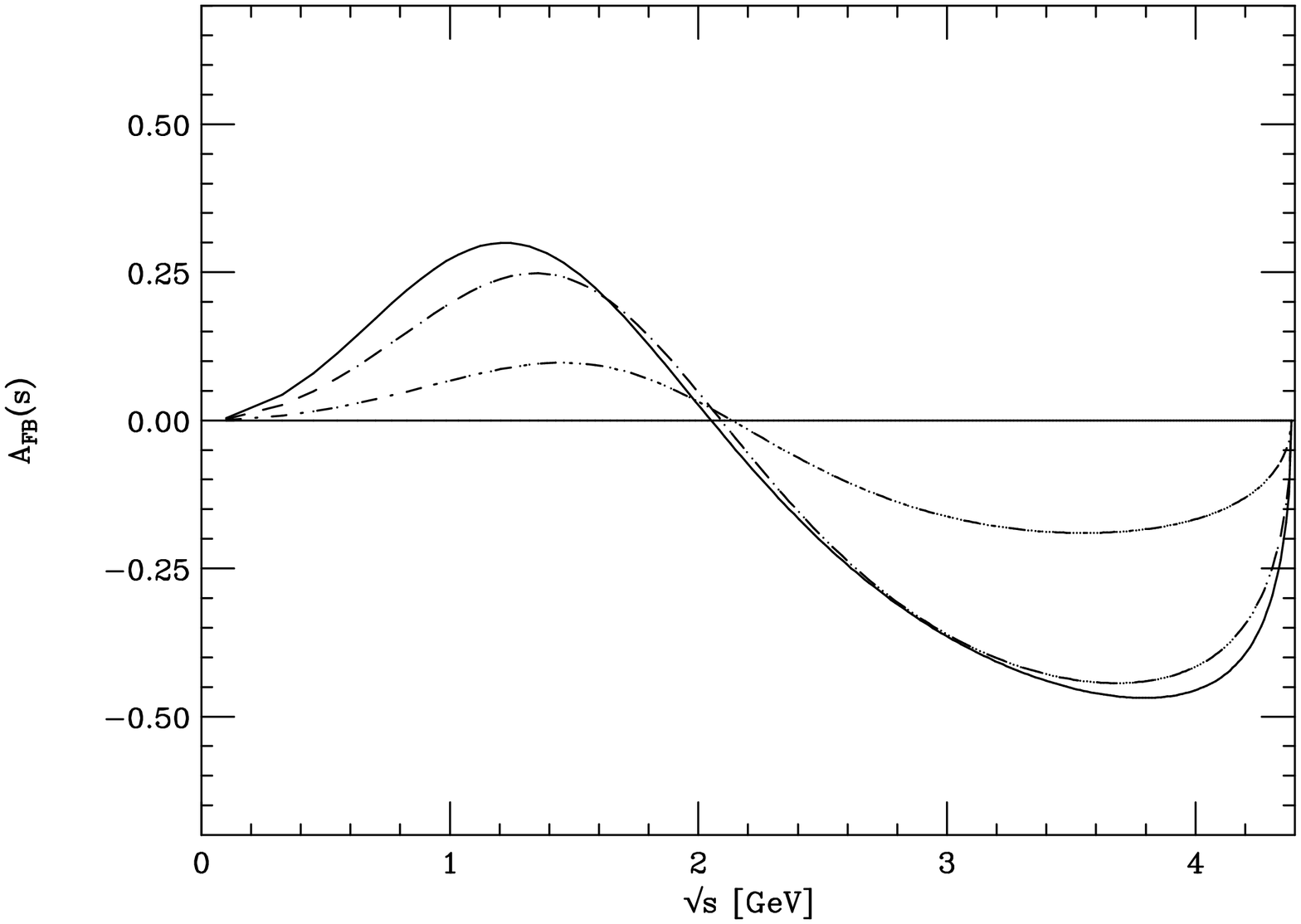,height=2.5in,angle=90}
}
\caption{ a) {\em The $Br(b\to s\ell^+\ell^-)$, normalized to the SM expectation, 
vs. $\Delta g_1^Z$ (solid) and $g_5^Z$ (dashed).} 
b) {\em The forward-backward lepton asymmetry in $B\to K^*\ell^+\ell^-$ vs. the 
dilepton mass $\sqrt{s}$, for $\Delta g_1^Z=0$,$~0.1$ and $0.2$ 
(solid, dashed, dot-dashed respectively). From Ref.~[21].}
}
\label{fbsll}
\end{figure}

Thus, at the energies at hand in rare decays, the only relevant coefficients are 
the three CP conserving parameters ($\Delta\kappa_\gamma$,~$\Delta g^Z_1$,~$g_5^Z$); 
and a CP violating one, $\tilde\kappa_\gamma$.  
Their FCNC effects are most interesting in $B$ and $K$ decays. For instance, in 
Fig.~\ref{fbsg}a we see the sensitivity of the current measurements of the 
$b\to s\gamma$
rate to the presence of $\Delta\kappa_\gamma$, whereas in Fig.~\ref{fbsg}b the 
branching ratio for $K^+\to\pi^+\nu\bar\nu$, normalized to the SM, is plotted against 
both $\Delta g_1^Z$ and $g_5^Z$.   
The effect of $\Delta\kappa_\gamma$ in $b\to s\gamma$ is obtained without any 
assumption 
other than the suppression of $\lambda_\gamma$. Similarly, from Fig.~\ref{fbsg}b
we see that the only significant
contribution of anomalous TGC to $K^+\to\pi^+\nu\bar\nu$ is given by $\Delta g_1^Z$, 
which could give effects as large as factors of $(2-3)$ in the branching ratio.
The CP violating parameter $\tilde\kappa_\gamma$ gives smaller effects that 
its CP conserving
counterpart since it does not interfere with the SM.  
In $b\to s\ell^+\ell^-$ decays again these two coefficients ($\Delta\kappa_\gamma$,
$\Delta g_1^Z$) give the dominant contributions. In Fig.~\ref{fbsll}a we see the effects
of the $WWZ$ couplings on the SM normalized branching ratio, taking
$\Delta\kappa_\gamma=0$. Somewhat less dramatic effects are given by this $WW\gamma$ coupling
by itself. However, an interesting feature of these decay modes, is that the additional 
information given by the lepton asymmetry can be used to disentangle the two contributions.
As it can be seen in Fig.~\ref{fbsll}b, the effect of $\Delta g_1^Z$ on the 
forward-backward lepton asymmetry $A_{\rm FB}(s)$ in $B\to K^*\ell^+\ell^-$ 
is such that it does not change the 
position of the zero~\cite{tgc}, but affects the shape of the asymmetry as well as the rate. 
On the other hand, the zero of $A_{\rm FB}(s)$ is shifted significantly by accessible 
values of $\Delta\kappa_\gamma$. Thus, a measurement of this asymmetry, as well as the 
rate, provides enough information to constrain the two relevant 
anomalous TGC without assuming that one of them vanishes. 
Then, the bounds obtained from FCNC decays 
are not only competitive with those from high energy colliders, but also complementary
to them due to their rather selective sensitivity. 
For comparison, we show in Table~III the projected sensitivities of 
LEPII~\cite{leptgc} at 
$\sqrt{s}=190$~GeV
and $500pb^{-1}$ integrated luminosity, the upgraded 
Tevatron~\cite{tevtgc} with $1fb^{-1}$, and a 
guess of the $3\sigma$ sensitivity to be reached in the next round of $B$ and 
$K$ experiments for FCNC decays. 
\phantom{xxxx}\vspace{0.1in}
\begin{center}
\begin{tabular}{|l||c|c|c|}
\hline
\multicolumn{4}{|c|}{Table~III.~{Comparison of bounds on Anomalous TGC. }}\\
\hline\hline
 & LEPII & Tevatron RunII & FCNC \\ 
 & 190~GeV & 1~fb$^{-1}$ & Decays\\ \hline
$\Delta\kappa_\gamma$ & (-0.25,0.40) & (-0.38,0.38) & (-0.20,0.20) \\ 
$\Delta g_1^Z$ & (-0.08,0.08) & (-0.18,0.48) & (-0.10,0.10) \\ 
$\tilde\kappa_\gamma$ & - & (-0.33,0.33) & (-0.50,0.50) \\ \hline
\end{tabular}
\end{center}
\phantom{xxxx}\vspace{0.1in}

\section*{Conclusions and Outlook}
Some of the FCNC decays we discussed are largely dominated by short distance physics, 
a fact that makes them very sensitive to extensions of the SM entering at one loop. 
This is particularly true of the $K\to\pi\nu\bar\nu$ modes as well as for the 
$B\to X_s\nu\bar\nu$ decays. The former are accessible at experiments planned for the 
near future, such as KAMI and CKM, whereas is not clear how to get SM sensitivity
for the neutrino modes in $B$ decays. 
On the other hand, $b\to s\gamma$ and $b\to s\ell^+\ell^-$ are short distance 
dominated modes. They may contain some long distance pollution 
as large as $20\%$, although this is theoretically very uncertain. 
At the moment this is not a limiting 
uncertainty, but it may become an issue when experiments such as LHC-B and BTeV 
start running.  

Rare charm decays are mostly dominated by long distance dynamics. This, 
in general, would prevents us from 
using this physics to test the short distance structure of the SM, but on the other hand
it constitutes a laboratory where we could improve our understanding of these effects, 
something we may need in rare $B$ decays. 
Also, there are some exceptions in charm physics, where one may still constrain 
considerably new physics scenarios. 
We mentioned the $c\to u\ell^+\ell^-$ modes, and also $D^0-\bar{D}^0$ mixing, 
have the potential to receive large non-standard effects that would appear somewhere 
between the current experimental limits and the most conservative estimates of the 
long distance effects. 

Finally, we have seen how rare FCNC decays complement the searches for new physics 
at high energy colliders. It is possible to imagine that, with the wealth of data on these 
physics that will be available in the near future (BELLE, BaBar, $K$ and 
$B$ physics at the Tevatron main injector, BNL and CERN $K$ experiments, etc.) 
a program similar to the  electroweak precision measurements of the 1990's could   
emerge. This program would then serve as guidance to the high $p_T$ physics to be 
carried out at the Tevatron, the LHC and beyond.

\vskip0.5cm
\noindent
{\bf Acknowledgments}

\noindent
This work was supported  by the U.S.~Department of Energy 
under  
Grant No.~DE-FG02-95ER40896 and the University of 
Wisconsin Research Committee with funds granted by the Wisconsin 
Alumni Research Foundation.

\end{document}